\documentclass[reprint,aps,superscriptaddress]{revtex4-2}
\usepackage{euscript}
\usepackage{amssymb}
\usepackage{amsfonts}
\usepackage{amsbsy}
\usepackage{epsfig}
\usepackage{subfigure}
\usepackage{amsthm}
\usepackage{mathrsfs}
\usepackage{amscd}
\usepackage{amstext}
\usepackage{verbatim}
\usepackage{amsmath}
\usepackage{url}
\usepackage{bm}
\usepackage{hyperref}
\usepackage{color}

\begin{document}
\title{Transition of vortex dipole dynamics in holographic superfluids}

\author{Yu-Kun Yan}
\email{yukunyan@zju.edu.cn}
\thanks{Corresponding author}
\affiliation{School of Physical Sciences, University of Chinese Academy of Sciences, Beijing 100049, China}
\affiliation{School of Aeronautics and Astronautics, Zhejiang University, Hangzhou, 310027, China}
\affiliation{Huanjiang Laboratory, Zhejiang University, Zhuji 311899, China}
\affiliation{National Key Laboratory of Aerospace Physics in Fluids, Mianyang 621000, China}
\author{Shanquan Lan}
\email{lansq@lingnan.edu.cn}
\thanks{Corresponding author}
\affiliation{Department of Physics, Lingnan Normal University, Zhanjiang 524048, China}
\author{Yu Tian}
\email{ytian@ucas.ac.cn}
\thanks{Corresponding author}
\affiliation{School of Physical Sciences, University of Chinese Academy of Sciences, Beijing 100049, China}
\author{Hongbao Zhang}
\email{hongbaozhang@bnu.edu.cn}
\thanks{Corresponding author}
\affiliation{School of Physics and Astronomy, Beijing Normal University, Beijing 100875, China}
\affiliation{Key Laboratory of Multiscale Spin Physics, Ministry of Education, Beijing Normal University, Beijing 100875, China}

\begin{abstract}
Using holographic duality, we reveal a transition in vortex dipole dynamics below a critical dipole size in strongly interacting superfluids, characterized by a significant suppression of mutual friction. In the bulk, this transition is triggered by a topological reconnection of vortex tubes, which disconnects the boundary vortices from the black hole horizon and forms a \textit{U-pipe}. Consequently, the post-transition evolution is governed by the contraction of the bulk \textit{U-pipe} rather than the mutual friction associated with the horizon, revealing a scale-dependent dissipation mechanism.
We further show that this reconnection persists over a broad temperature range, even when the transition becomes unobservable at high temperatures.
Our results provide a dissipation-based interpretation for the anomalous critical dipole scale observed in strongly interacting cold-atom experiments, and suggest the existence of distinct dissipative regimes in strongly interacting superfluids.
\end{abstract}

\maketitle

\emph{Introduction}---The superfluid dissipation is fundamentally tied to the dynamics of quantized vortices, whose motion governs a wide range of non-equilibrium phenomena including the resistive behavior of superconductors as well as quantum turbulence \cite{Makoto:2013}.
In weakly interacting Bose condensates, the vortex dynamics is well described by the dissipative Gross-Pitaevskii equation (DGPE), where the dissipation is incorporated phenomenologically to capture the mutual friction between the vortices and normal fluid \cite{Hall:1956b}.
However, the dissipation mechanism in strongly interacting superfluids remains much less understood, since the microscopic origin of dissipation is difficult to access beyond the quasi-particle descriptions \cite{Liu:2020,Sachdev:2011phystoday,Hartnoll:2018xxg}. 
This issue has become particularly intriguing in recent cold-atom experiments near unitarity, where vortex dipoles exhibit an anomalous critical dipole scale (about 10 vortex radii) during the vortex dipole shrinkage, below which the dipole seems rapidly annihilate \cite{Kwon:2021}.
Such critical dipole size is
much larger than that in the weakly interacting case
which is approximately equal to the healing length \cite{Jones1982,Janne:2005}.
A related counterintuitive dissipation behavior has also been observed near unitarity and on the BCS side, where vortices decay faster at lower temperatures \cite{Liu2022}. Although these anomalous dissipative behaviors have been conjectured to originate from the weaker pairing strength and the presence of unpaired fermions, a first-principle understanding of the underlying dissipation mechanism remains lacking.

Gauge/gravity duality (holography) provides a non-perturbative framework for studying strongly interacting quantum systems through higher-dimensional classical gravity \cite{tHooft:1993,Susskind:1994,Maldacena:1997,Witten:1998}.
In this framework, the finite temperature dissipation is geometrically encoded by the black hole horizon, which equals to the presence of a thermal bath for the boundary quantum system \cite{Sonner2012,Hartnoll:2018xxg,Liu:2020}. 
Benefiting from these properties, holography has been widely applied to dissipative and non-equilibrium dynamics of strongly correlated condensed matter systems \citep{Hartnoll:2008,Kovtun:2005,McGreevy:2009,Solana_Liu:2023,Zaanen:2015}, including the defect formation \cite{Chesler:2015s,Sonner:2014tca,Yang2025}, dynamical transitions \cite{Guo2020,Lan:2023}, and time translation symmetry breaking \cite{Yang:2023}.
Among these, holography has also demonstrated its capability in capturing the dissipation properties of quantum turbulence from the statistical perspective \cite{Adams:2012,Yang:2024,Zeng:2024}, and also the vortex motion from the microscopic perspective \cite{Lan:2019,Ewerz:2021,Wittmer:2021}.
However, how vortex motion varies with temperature and spatial scale remains unclear, especially during the late-stage evolution preceding vortex annihilation which is crucial for understanding the experimental observations.
\begin{figure*}[tbp]
\centering
  \includegraphics[width=17cm]{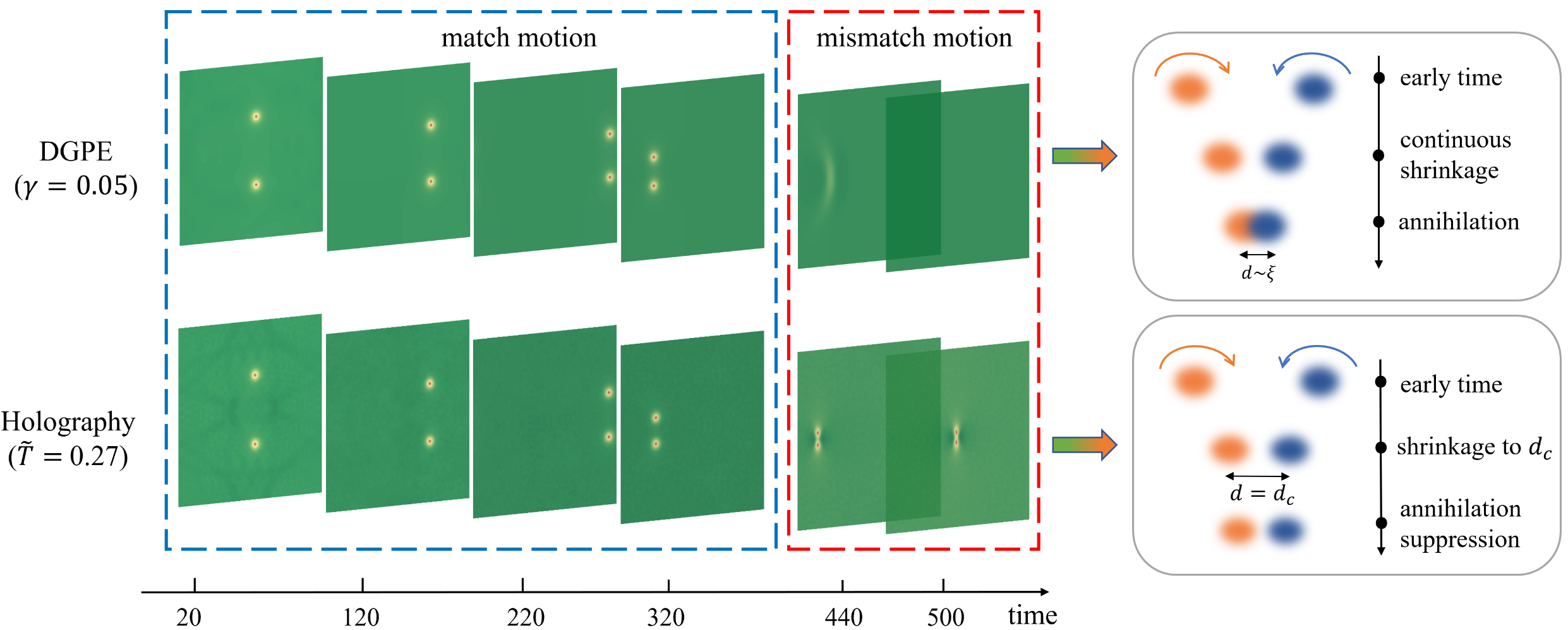}
\caption{The comparison of vortex dipole motion in the DGPE and holographic superfluids. The snaps of condensate density $|\psi|^2$ in the blue box denotes the motion that can be matched with $\gamma=0.05$, $\tau=13.3$ and $\mu=16.9$, while the red box shows the mismatch motion. The schematic diagrams of both cases are shown on the right side. In DGPE, the vortex and anti-vortex will get close to each other until $d\sim \xi$ and annihilate. While in holographic superfluids, the dipole will undergo the same process as DGPE until $d\sim d_c$, and then the annihilation is suppressed apparently.}\label{fig1}
\end{figure*}

In this work, we systematically investigate the vortex dipole dynamics utilizing holographic superfluids, and uncover a transition in vortex dipole dynamics below a critical dipole size, which cannot be described by DGPE with a single dissipation parameter, offering a dissipation-based explanation for the critical size observed in \cite{Kwon:2021}.
The bulk fields evolution suggests that this transition is triggered by a topological reconnection of bulk vortex tubes, which indicates a
scale-dependent dissipation mechanism in strongly interacting superfluids.
This topological reconnection remains robust against temperature variation.
The dissipation behavior across the critical dipole size is further tested through the friction coefficient, dissipation rate and sound emission, which supports the scale-dependent nature of the dissipation mechanism.

\emph{Holographic superfluids and DGPE}---
From the lens of gravity, the strongly interacting superfluids can be effectively described by the Abelian Higgs model in an asymptotically $AdS$ black hole background, which has been first implemented in \cite{Hartnoll:2008}. In the probe limit, the Lagrangian density takes the form
\begin{eqnarray}
\mathcal{L}&=&-\frac{1}{4}F^{\mu\nu}F_{\mu\nu}-|D_\mu\Phi|^2-m^2|\Phi|^2,
\end{eqnarray}
where $F_{\mu\nu}=\nabla_\mu A_\nu-\nabla_\nu A_\mu$ and $D_\mu=\nabla_\mu-iA_\mu$. We consider the theory defined in the planar Schwarzschild $AdS_{3+1}$ black hole background with Eddington-Finkelstein coordinates,
\begin{equation}
ds^2=\frac{L_{AdS}^2}{z^2}(-f(z)dt^2+dx^2+dy^2-2dtdz),
\end{equation}
where $L_{AdS}$ is the AdS radius and $f(z)=1-(z/z_h)^3$ with $z_h$ the location of the black hole horizon. 
The equations of motion for the matter fields read
\begin{equation}
\begin{split}
&\nabla_{\mu}F^{\mu\nu}=J^\nu=i(\Phi^*D^{\nu}\Phi-\Phi(D^{\nu}\Phi)^*),\\
&D_{\mu}D^\mu\Phi-m^2\Phi=0,
\end{split}\label{eoms}
\end{equation}
where the asterisk denotes the complex conjugation.
The boundary system temperature is given by Hawking temperature $T_H=3/(4\pi z_h)$. The chemical potential $\tilde{\mu}$ read as $A_t|_{z=0}$ and $-\partial_z A_t|_{z=0}=\rho$ the particle number density.
For convenience, we set $L_{AdS}=z_h=1$ and choose $m^2=-2$ without loss of generality.
When $\tilde{\mu}>\tilde{\mu}_c$ with $\tilde{\mu}_c=4.064$ the critical chemical potential, the Higgs mechanism will be triggered and the bulk complex scalar field will spontaneously condense, which signals the superfluid transition on the boundary, and the order parameter $\psi$ can be read from the scalar field $\Phi$.
Since the anomalous vortex dynamics is observed near unitarity and on the BCS side in fermionic superfluids, we focus on the standard quantization of the holographic superfluid model, which is qualitatively associated with the strongly interacting BCS regime \cite{Ville2010a, Ville2010b}.
We define the dimensionless temperature as $T=T_H/\sqrt{\rho}$ and the superfluid transition temperature $T_c=T_H/\sqrt{\rho_c}$.
Then a reduced temperature is given by $\tilde{T}=T/T_c$.

On the other hand, for weakly interacting bosonic superfluids such as Bose-Einstein condensates (BECs), the finite temperature dynamics can be well described by the DGPE, which can be written as
\begin{equation}
\partial_t \psi=\tau(i+\gamma)\mathcal{L}_{\text{gp}},
\end{equation}
where $\mathcal{L}_{\text{gp}}=\nabla^2\psi/2+\mu(1-|\psi|^2)\psi$ and $\tau$ is an adjustable time coefficient.
$\mu$ is the chemical potential of condensate and the healing length is defined as $\xi=1/\sqrt{2\mu}$. $\gamma$ is the dissipation parameter which can be determined by the Keldysh self-energy through the fluctuation-dissipation theorem \cite{Stoof:1997,Stoof:1999,Duine:2001}. It describes collisions between the normal reservoir atoms and atoms in the BEC, which has been widely used in the vortex and quantum turbulence dynamics \cite{Bradley:2012, Reeves:2013}.

\emph{The transition of vortex dipole motion}---For revealing the full vortex dipole dynamics, we perform the time evolution of the vortex dipole in a periodic box of length $L=2 r_v n_L$ where $r_v$ is the vortex radius and $n_L$ is the dimensionless size of the periodic box, so $n_L^2$ representing the maximum number of vortices the box can accommodate. The initial dimensionless dipole size is set as $d(t=0)/r_v=4n_L/5$.
Choosing $\tilde{T}=0.27$, we compare the dipole evolution in holography with that in DGPE by matching the dissipation parameter $\gamma=0.05$ as shown in Fig.(\ref{fig1}).
It displays that holography and DGPE can have the same vortex dynamical trajectories when the dipole size $d$ is large which has also been shown in \cite{Wittmer:2021,Yan2023}.
However, when the vortices approach each other, the dipole motion can no longer be matched
by $\gamma=0.05$.
In the DGPE, the mutual friction will continue to act and lead the vortices annihilate when $d\sim \xi$.
While in holography, a motion transition happens that the mutual friction between vortices and normal fluid seems to weaken suddenly when $d$ reaches to a critical size $d_c$, thereby suppressing the annihilation process.
The DGPE can not realize this behavior with a single dissipation parameter, which indicates the scale dependence of the dissipation mechanism in the BCS-like strongly interacting superfluids.

\begin{figure}[tbp]
\centering
\subfigure[]{
  \includegraphics[width=4cm]{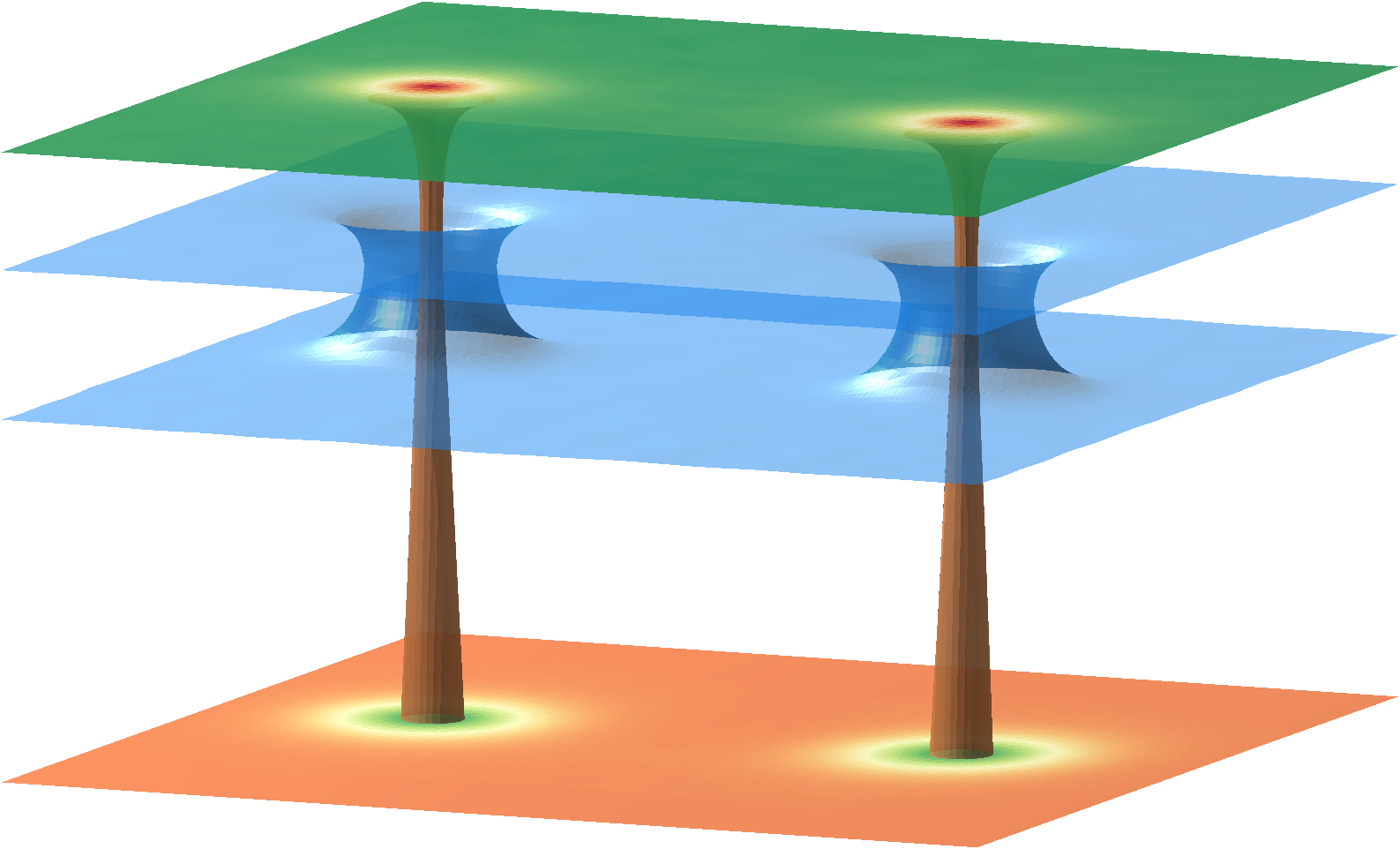}}
\subfigure[]{
  \includegraphics[width=4cm]{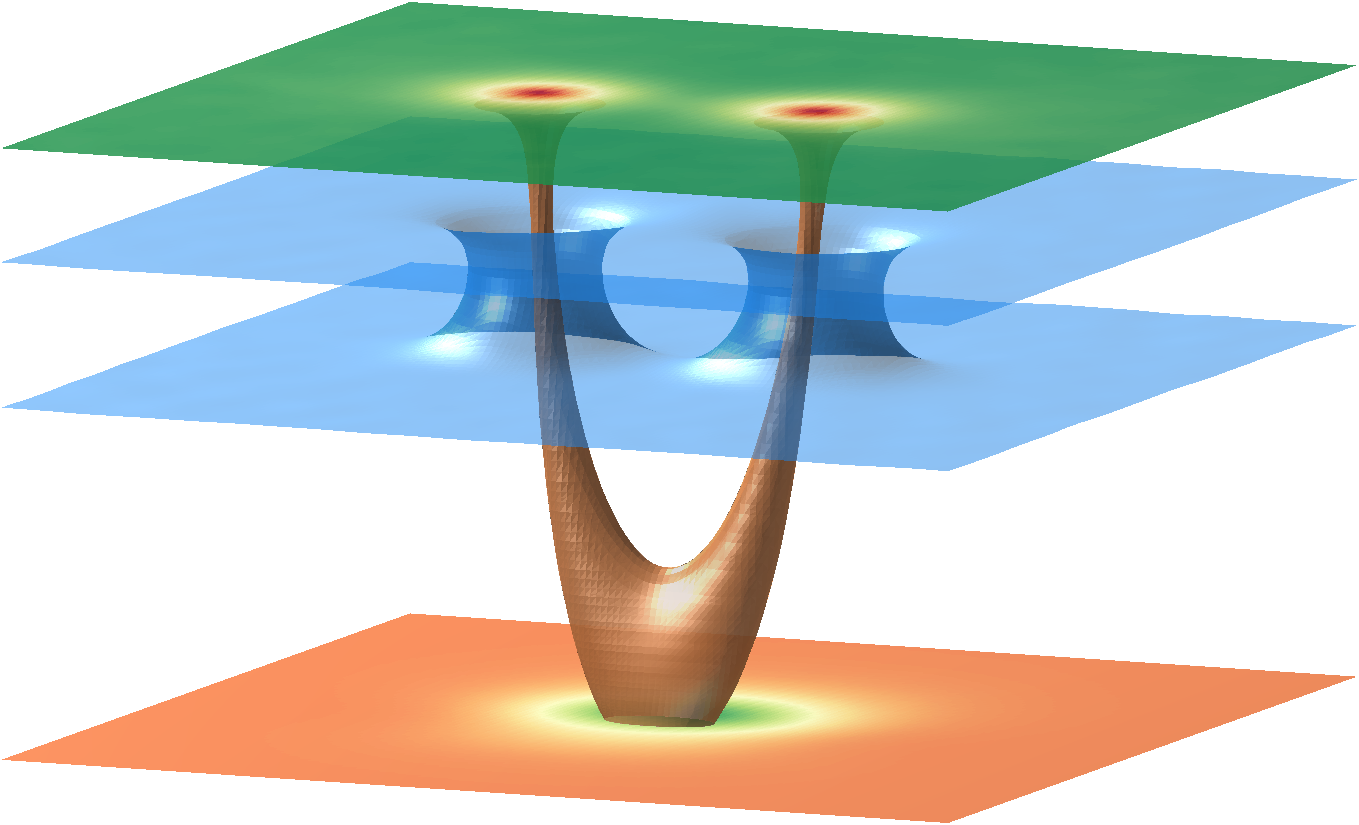}}
\subfigure[]{
  \includegraphics[width=4cm]{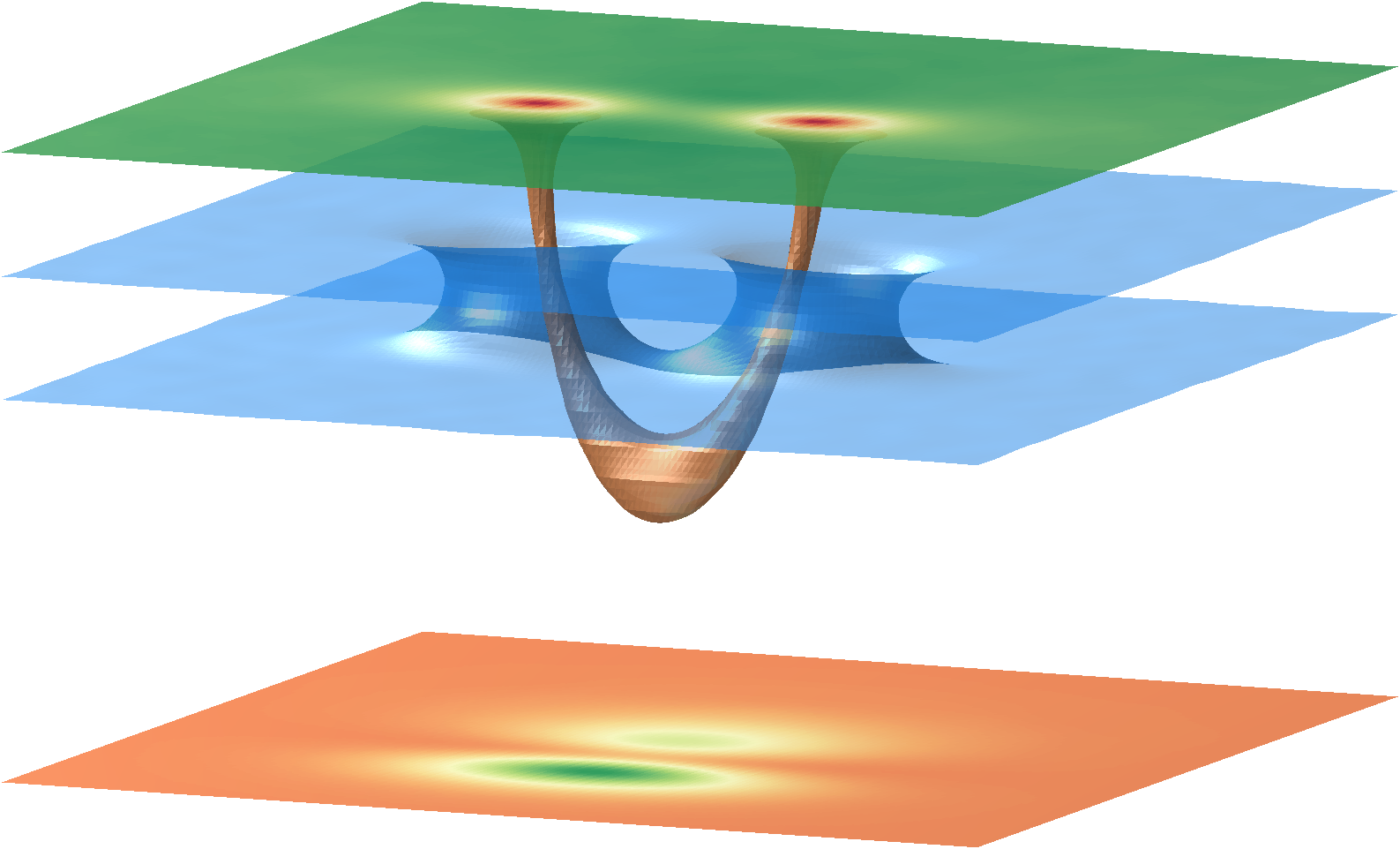}}
\subfigure[]{
  \includegraphics[width=4cm]{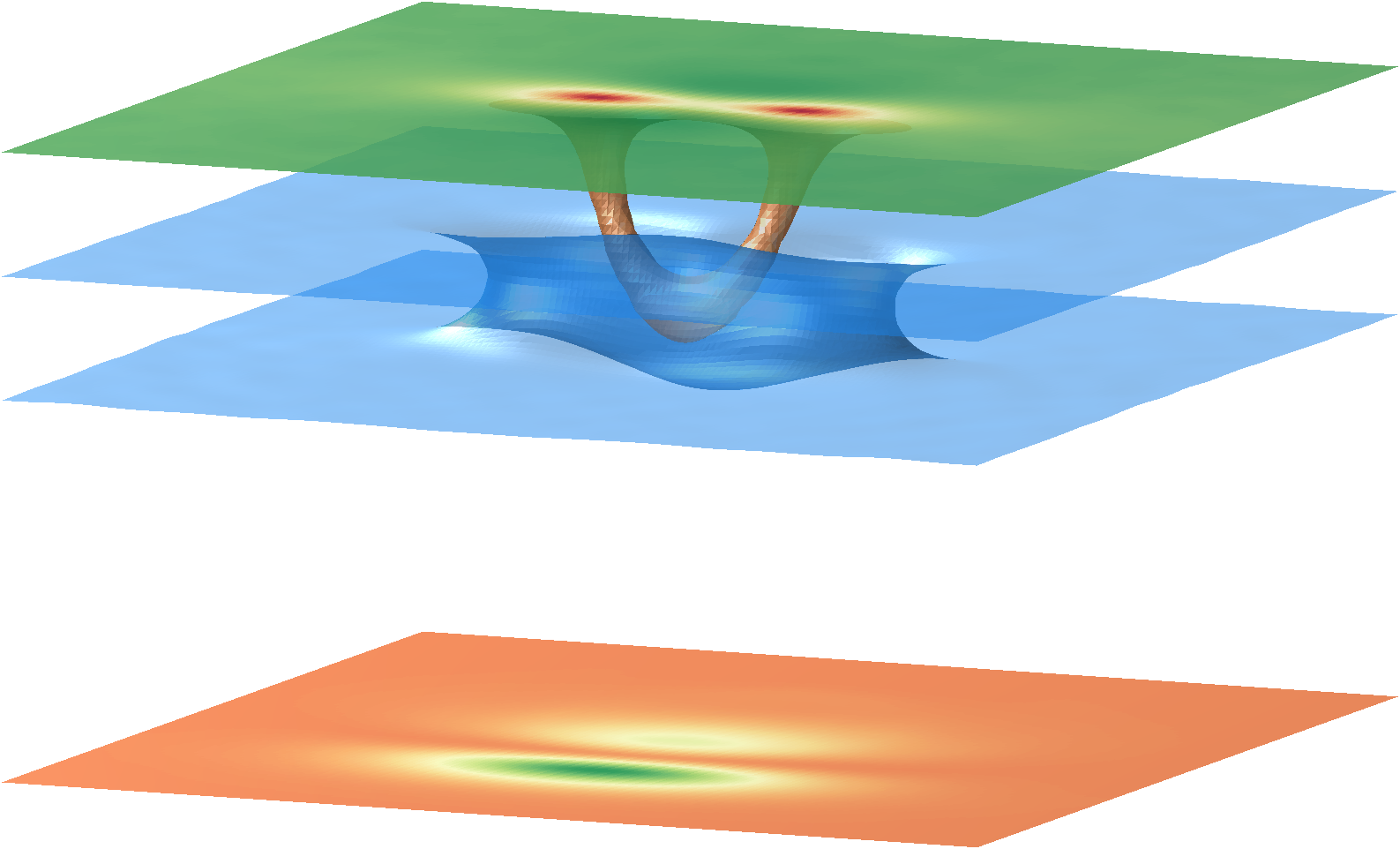}}
\caption{Bulk configurations of vortex dipole of four different evolution stages. The brown tubes are the iso-surface of $\Psi$. The  blue surface is the iso-surface of bulk charge density. The top surface is condensate $\psi$ and the bottom surface is energy flux at horizon $\tau^z_{\ t}|_{\text{horizon}}$ . (a) is before the transition. (b) is when the transition happens. (c-d)  shows the vortex shrinkage after the transition and contraction of \textit{U-pipe}.}\label{fig2}
\end{figure}

\begin{figure}[tbp]
\centering
\subfigure{
  \includegraphics[width=8cm]{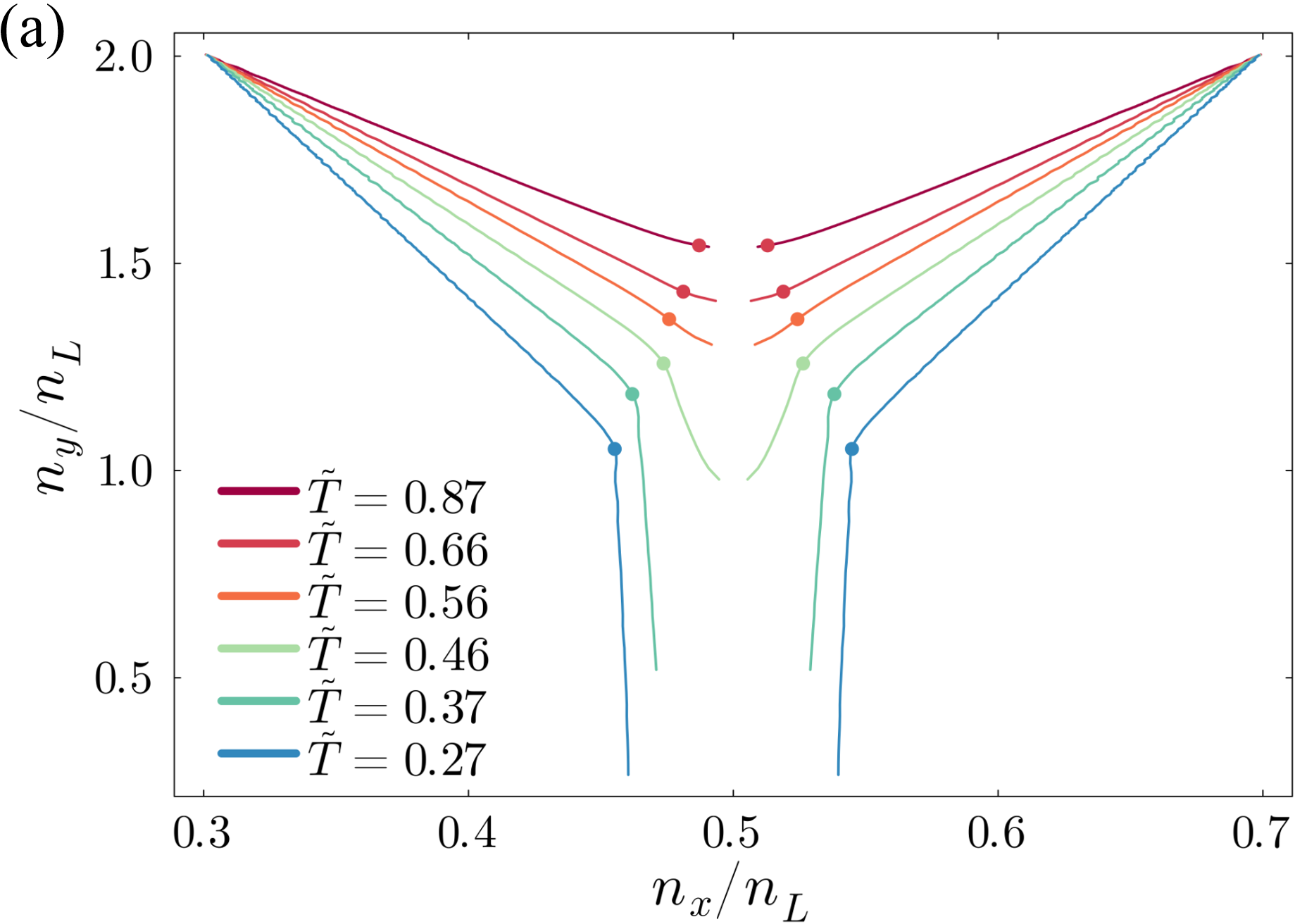}}
\subfigure{
  \includegraphics[width=8cm]{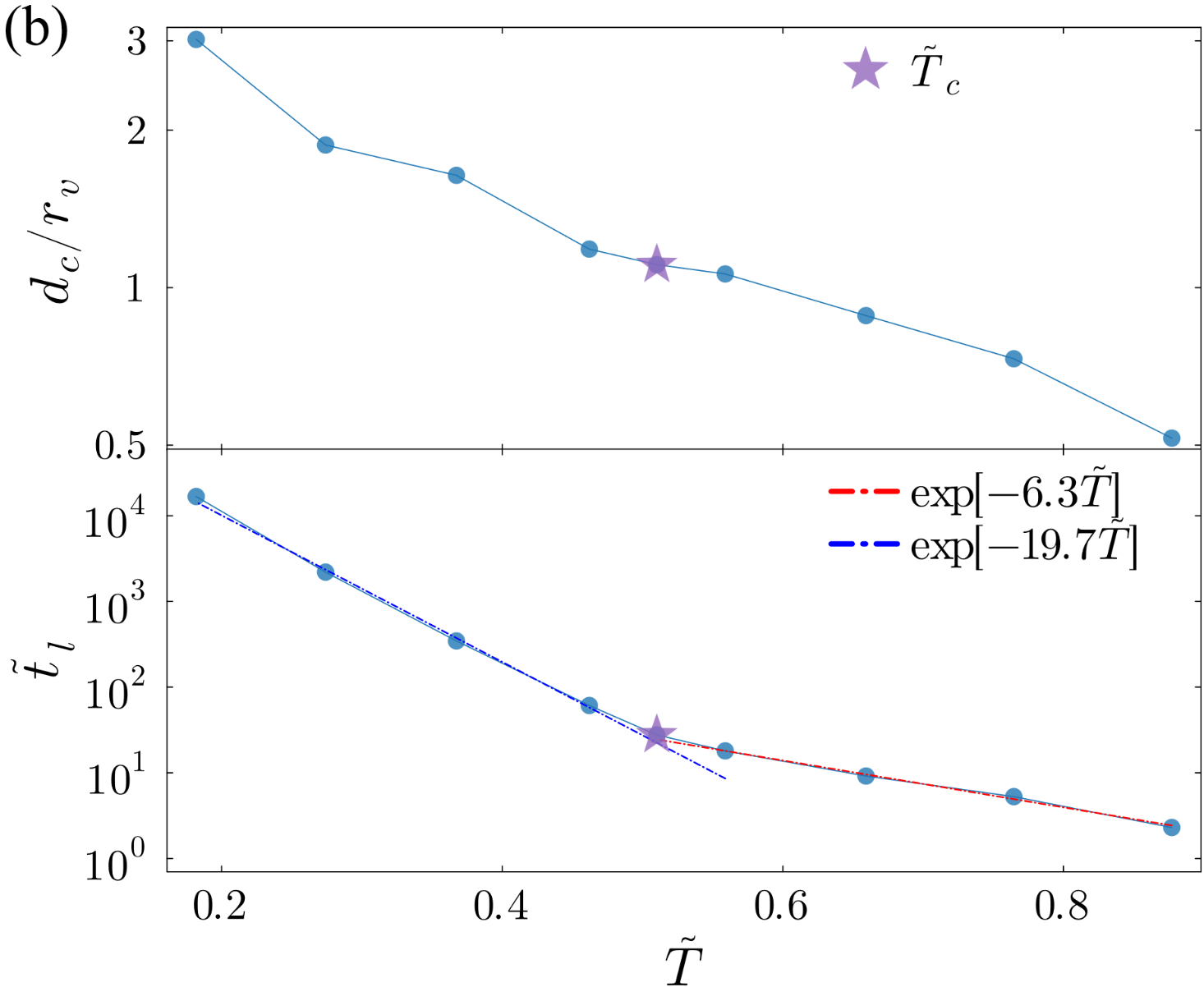}}
\caption{The temperature dependence of the dipole motion.(a) shows the vortex dipole trajectories under different temperatures, where $n_x$ and $n_y$ denotes the dimensionless positions of $x$ and $y$ respectively, and the dots label the vortex positions when transition happens. (b) displays the temperature dependence of the dimensionless critical dipole size $d_c/r_v$ and the lifetime of the \textit{U-pipe} $\tilde{t}_l$. The purple star labels the characteristic temperature $\tilde{T}_c\approx0.51$. The dimensionless time $\tilde{t}$ is defined as $t\sqrt{\rho}$.}\label{fig3}
\end{figure}

As previously mentioned that the holographic duality provides a first-principle dissipation description of the finite temperature boundary theory. In holographic superfluids model, this mechanism has been shown in \cite{Adams:2014} that the vortex in the boundary duals to a flux tube in the bulk, through which the energy perturbations near the boundary pass to the horizon unimpeded, thereby realizing the mutual friction of vortex and the normal fluids.
To understand the behavior of scale dependence and $d_c$, we show such bulk configuration of vortex dipole in Fig.(\ref{fig2}).
The blue slab represents the bulk charge distribution, which screens boundary excitations from directly falling into the horizon, while the vortices punch holes through this screening slab, providing low-charge channels for excitations to fall into the black hole \cite{Adams:2014}.
Before the transition, the boundary vortex dipole is connected to the horizon through straight flux tubes, leading to dynamically consistent dipole trajectories between holography and DGPE.
However, once the boundary motion transition occurs, the vortices at the horizon annihilate in advance, and the bulk flux tubes undergo a topological reconnection forming a \textit{U-pipe} structure as shown in Fig.(\ref{fig2}b).
This reconnection breaks the direct dissipative connection between the boundary vortices and horizon.
As a result, this transition reflects a change in the dominant dissipation mechanism across $d_c$: from mutual friction associated with the horizon to the bulk dynamics associated with the U-pipe
contraction as shown in Fig.(\ref{fig2}b-d) which is more sensitive to the short-distance physics.
Meanwhile, the critical dipole size $d_c$ of the transition can be identified as the vortex dipole size at boundary when the horizon dipole annihilates.

\emph{The temperature dependence of vortex dipole motion}---
We further investigate the effect of temperature on the vortex dipole dynamics. Fig.(\ref{fig3}a) illustrates trajectories of dipole under varying reduced temperatures $\tilde{T}$, where dots mark the vortex positions at the critical dipole size $d(t)=d_c$.
As shown in Fig.(\ref{fig3}b), the dimensionless critical dipole size $d_c/r_v$ remains finite even as $\tilde{T} \rightarrow 1$, suggesting that the \textit{U-pipe} topological reconnection consistently precedes self-annihilation within the explored temperature range.
However, at relatively high temperatures, the transition becomes indiscernible in the trajectories, because $d_c$ drops below the vortex radius $d_c<r_c\sim\xi$.
Furthermore, the dimensionless lifetime $\tilde{t}_l$ of the \textit{U-pipe}, defined as the duration between the transition and the final annihilation, also exhibits two distinct exponential decay rates with temperature in Fig.(\ref{fig3}b).
These allow us to identify a characteristic temperature $\tilde{T}_c \approx 0.51$, corresponding to $d_c/r_v \approx 1$.
Consequently, the motion transition becomes clearly observable only when $\tilde{T} < \tilde{T}_c$ (where $d_c/r_v > 1$).
Due to numerical limitations at very low temperatures, our analysis is restricted to $\tilde{T}\gtrsim0.18$, and the fate of the transition in the near zero temperature limit remains an open question.

\begin{figure}[tbp]
\centering
\subfigure{
  \includegraphics[width=9cm]{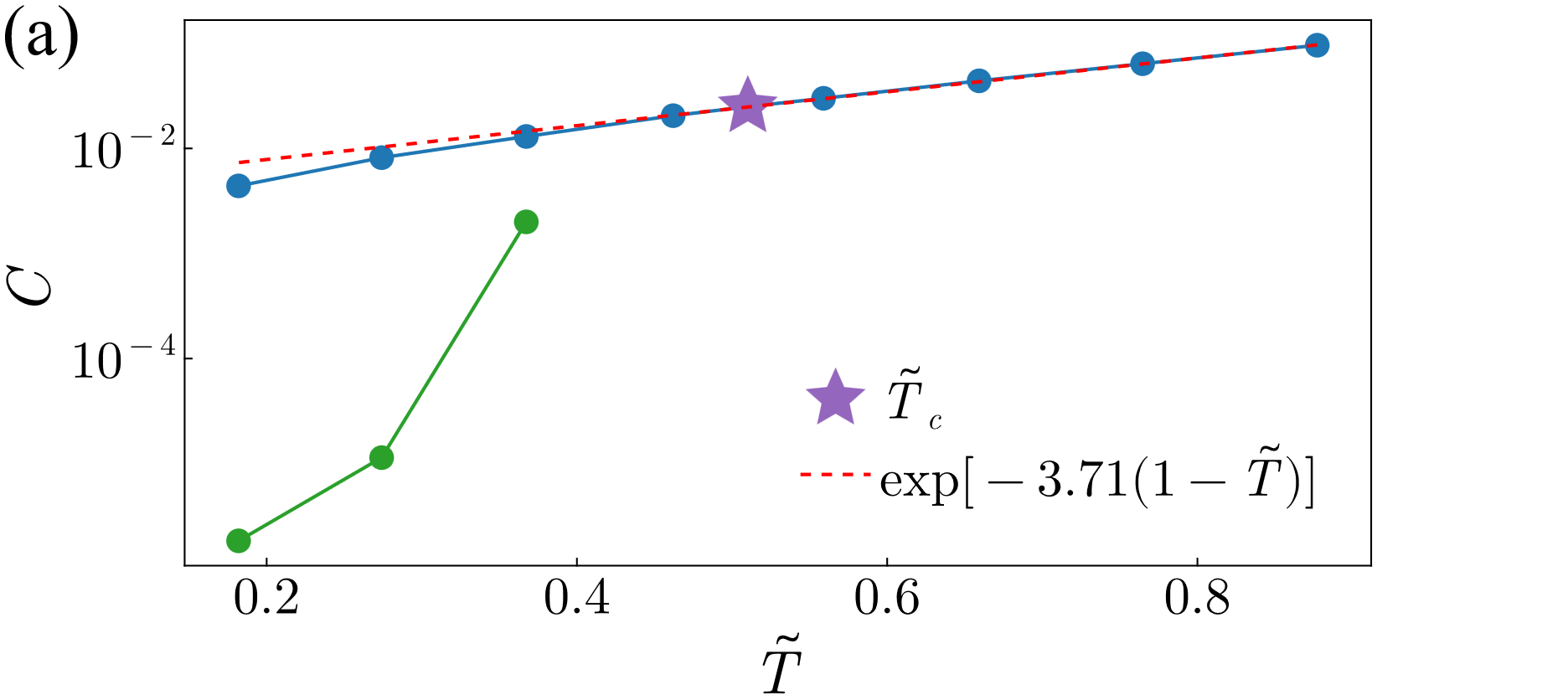}}
\subfigure{\includegraphics[width=9cm]{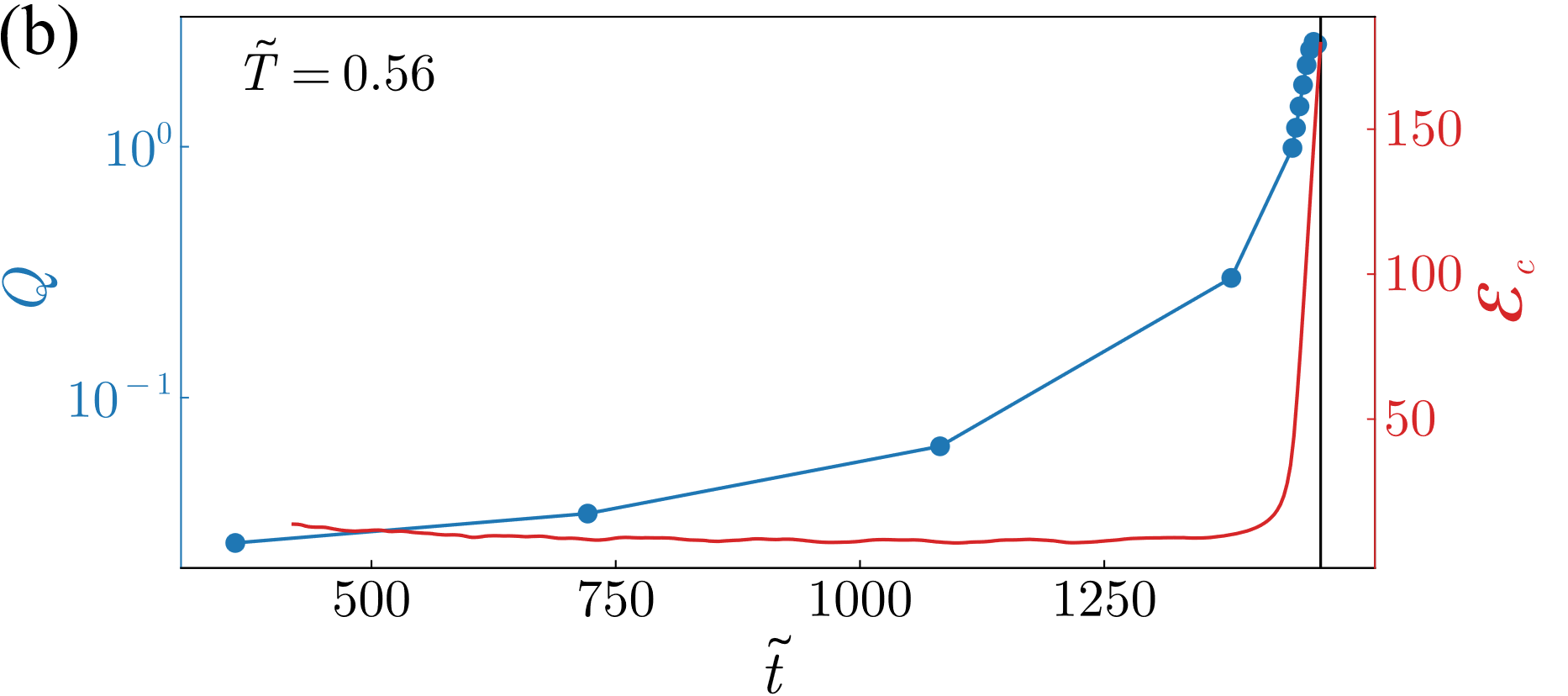}}
\subfigure{\includegraphics[width=9cm]{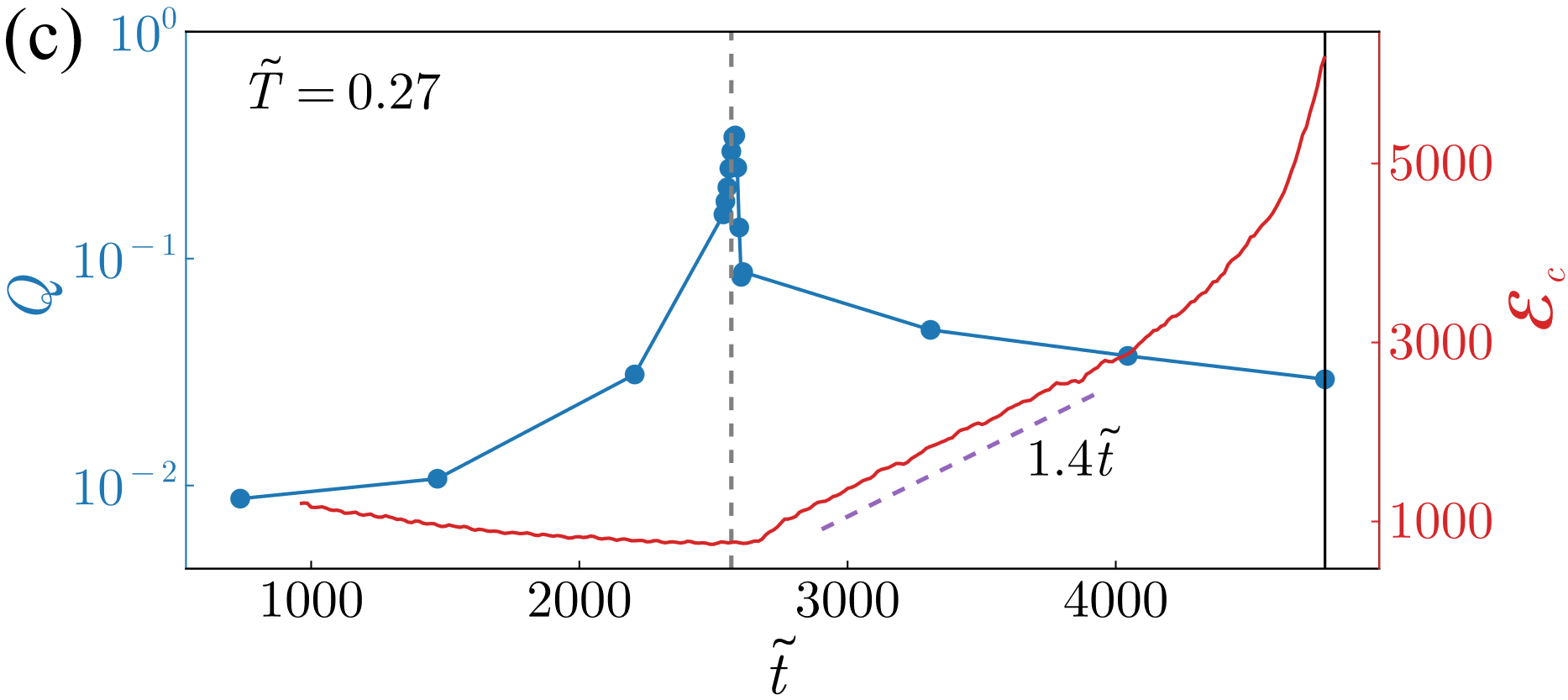}}
\caption{The dissipative properties across the transition. (a) shows the temperature variation of the friction coefficient $C$. The green dot line labels $C$ of post-transition and the red dash line labels the exponential decay behavior. (b-c) displays the variations of total dissipation rate $Q$ (blue dot line) and compressible kinetic energy $\mathcal{E}_c$ (red line) when $\tilde{T}=0.56$ and $\tilde{T}=0.27$ respectively. The gray dash line labels the time when the transition happens and the black line labels the time when vortices annihilate.}\label{fig4}
\end{figure}

\emph{The dissipation behavior of vortex dipole motion}---
To characterize the dissipative properties across the transition, we further analyze the effective mutual friction, horizon energy dissipation, and sound emission.
At the boundary, the finite-temperature mutual friction can be quantitatively described by the point-vortex model \cite{HVI:1956,HVI:1964}. For sufficiently large dipole sizes, the friction coefficient $C$ can be extracted from $d(t)=\sqrt{d(0)^2-8Ct}$.
The results under varying $\tilde{T}$ are shown in Fig.(\ref{fig4}a). 
At relatively high temperatures, where the dynamical transition becomes unobservable, the friction coefficient can be uniquely extracted and exhibits an exponential decay $C\propto \textrm{exp}[-3.71(1-\tilde{T})]$.
When $\tilde{T}<\tilde{T}_c$, the transition becomes pronounced, and two distinct friction coefficients can be identified before and after the transition for $\tilde{T}\lesssim 0.4$ as indicated by the branching behavior of $C$.
In the pre-transition regime (blue dots), $C$ decreases faster than the exponential behavior at high temperatures, but remaining within $\mathcal{O}(10^{-3})\sim\mathcal{O}(10^{-1})$. 
By contrast, in the post-transition regime (green dots), $C$ exhibits a qualitatively different temperature dependence and becomes several orders of magnitude smaller than the pre-transition value as $\tilde{T}$ approaches to zero.
From the holographic perspective, the finite temperature dissipation is characterized by the energy flux at horizon $\tau^z_{\ t}|_{\text{horizon}}$,
and the total dissipation rate is given by $Q=-\int dx^2 \tau^z_{\ t}|_{\text{horizon}}$ \cite{Adams:2012, Zeng:2024}.
For the conventional vortex dipole shrinkage and annihilation, $Q$ will increase as the dipole size decreases, as shown in Fig.(\ref{fig4}b) for $\tilde{T}=0.56$ where the dynamical transition is unobservable. 
In contrast, when the transition becomes pronounced at $\tilde{T}=0.27$, Fig.(\ref{fig4}c) shows that the dissipation rate undergoes a sudden decrease as the \textit{U-pipe} forms, followed by a much milder evolution until the final annihilation.

In addition to the finite temperature dissipation, there is also the sound emission that can translate the incompressible kinetic energy into compressible kinetic energy when the dipole is about to annihilate.
This is shown in Fig.(\ref{fig4}b) as $\tilde{T}=0.56$, where the compressible kinetic energy $\mathcal{E}_c$ experiences a rapid increase when the dipole is annihilating.
Remarkably, a distinct sound-emission behavior is observed when $\tilde{T}=0.27$: $\mathcal{E}^c$ shows a linear increase with time after the motion transition, demonstrating that the formation of $\textit{U-pipe}$ not only affects the finite temperature dissipation, but also turns the dipole into a long-lived sound source.
Together, these results further demonstrate that the dissipation mechanism changes qualitatively across the transition scale $d_c$.

\emph{Conclusions}---In summary, by comparing vortex dipole dynamics in holographic superfluids and the dissipative Gross-Pitaevskii equation, we uncover a transition of the dissipative process at a critical dipole size $d_c$.
Through holography, this transition admits a geometric interpretation: the bulk vortex tubes reconnect with each other and form a \textit{U-pipe}, which breaks the direct dissipative connection between the boundary vortices and the black hole horizon.
Consequently, the post-transition dynamics becomes governed by the contraction of the bulk \textit{U-pipe} rather than the conventional mutual friction. This transition is further characterized through the friction coefficient, the dissipation rate and sound emission.
Furthermore, the temperature dependence suggests that the transition persists over a broad temperature range $0.18\lesssim \tilde{T}<1$.
It would be important to investigate whether the observed transition survives in the near zero temperature limit, where the present numerics become increasingly challenging.

Our findings offer a dissipation-based explanation for the critical dipole size observed in the strongly interacting regime of cold-atom experiments \cite{Kwon:2021}, separating the dipole shrinkage from the subsequent annihilation.
However, while cold atom experiments indicate a rapid annihilation of vortex dipoles once their size falls below $d_c$, the detailed dynamics in this regime remain not yet fully resolved. Whether the post-transition evolution follows the trajectory structure suggested by the holographic framework thus remains an open question, motivating higher-precision experimental studies.
Indeed, the post-transition regime is expected to be more sensitive to the short-distance physics of the system, and therefore may exhibit model-dependent dynamics beyond the \textit{U-pipe} behavior captured holographically.

\emph{Acknowledgments}---We gratefully acknowledge the insightful discussions with Hong Liu and Woojin Kwon. Y.\ K.\ is supported by the Research Fund of National Key Laboratory of Aerospace Physics in Fluids with Grants No.\ 2025-APF-KFQMJJ-11. S.\ L.\ is  supported by the funding of National Natural Science
Foundation of China with Grants No.\ 12575057 and
Guangdong Basic and Applied Basic Research Foundation
of China with Grants No.\ 2024A1515012552 and No. 2026A1515012307. Y.\ T.\ is partially supported by the
National Natural Science Foundation of China with Grants No.\ 12375058.
H.\ Z.\ is partly supported by the National Natural Science Foundation of China with
Grants No.\ 12361141825 and No.\ 12575047.

\appendix
\section{Numerical details}

Having set $L_{AdS}=z_h=1$ and $m^2=-2$, together with the axial gauge $A_z=0$, the equations of motion in Eq.(\ref{eoms}) can be decomposed into the constraint equation
\begin{equation}
0=-\partial_z^2 A_t+\partial_z\partial\cdot\textbf{A}-2\textrm{Im}(\phi^*\partial_z\phi),\label{constrain}
\end{equation}
and the evolution equations
\begin{eqnarray}
\partial_t \partial_z \phi&=&\partial_z(\frac{f(z)}{2}\partial_z\phi)+\frac{1}{2}\partial^2\phi-i\textbf{A}
\cdot\partial\phi \label{eqphi}\\
&+&iA_t\partial_z\phi 
-\frac{i}{2}(\partial\cdot\textbf{A}-\partial_zA_t)\phi
-\frac{1}{2}(z+\textbf{A}^2)\phi, \notag \\
\partial_t \partial_z \textbf{A}&=&\partial_z(\frac{f(z)}{2}\partial_z\textbf{A})-|\phi|^2\textbf{A}+\textrm{Im}(\phi^*\partial\phi) \label{eqA}\\
&+&\frac{1}{2}\left[\partial\partial_zA_t+\partial^2\textbf{A} -\partial\partial \cdot\textbf{A}\right], \notag \\
\partial_t\partial_z A_t&=&\partial^2A_t-\partial_t\partial\cdot\textbf{A}+f(z)\partial_z\partial\cdot\textbf{A} -2A_t|\phi|^2 \label{eqAt}\\
&+&2\textrm{Im}(\phi^*\partial_t\phi)-2f(z)\textrm{Im}(\phi^*\partial_z\phi),
\notag
\end{eqnarray}
where $\phi=\Phi/z$ and $\textbf{A}=(A_x, A_y)$.
Near the AdS boundary $z=0$, the fields exhibit the asymptotic expansions
\begin{eqnarray}
\phi(t,\textbf{x},z)&=&s(t,\textbf{x})+\psi(t,\textbf{x}) z+\mathcal{O}(z^2),\\
A_\mu(t,\textbf{x},z)&=&a_\mu(t,\textbf{x})+j_\mu(t,\textbf{x}) z+\mathcal{O}(z^2),
\end{eqnarray}
with $\textbf{x}=(x,y)$. The boundary temperature is given by the Hawking temperature at horizon $T_H=3/(4\pi z_h)$, and the chemical potential $\mu$ is given by $a_t$. The total particle number density $\rho$ is identified as $-j_t$, and $a_\textbf{x}$ serves as the external source of the
current density $j_\textbf{x}$. Considering the standard quantization, $s(t,\textbf{x})$ plays the role of boundary source and $\psi(t,\textbf{x})$ is the vacuum expectation value of the scalar operator. When $\mu>\mu_c$ with $\mu_c=4.064$ the critical chemical potential, the Higgs mechanism will be triggered and the bulk complex scalar field will spontaneously condense, which signals the superfluid transition on the boundary, and the order parameter can be read from the  scalar operator $\psi$.

For the fully nonlinear numerical evolution, we impose the following boundary conditions:
\begin{eqnarray}
\phi|_{z=0}&=&0,\ \ \textbf{A}|_{z=0}=0, \label{cd1}\\
A_t|_{z=0}&=&\mu,\ \ \partial_z A_t|_{z=0}=-\rho(t,\textbf{x}).\label{cd2}
\end{eqnarray}

We evolve $\phi$ and $\textbf{A}$ by Eq.(\ref{eqphi}) and Eq.(\ref{eqA}) with Eq.(\ref{cd1}). $A_t$ is solved by the constraint equation with Eq.(\ref{cd2}). The dynamic of $\rho$ is given by Eq.(\ref{eqAt}) evaluated at the boundary, which is simply the current conservation law. Owing to the constraint equation, once Eq.(\ref{eqAt})is satisfied at the boundary, it will hold in the whole bulk. Periodic boundary conditions are imposed in the $\textbf{x}$ directions for all fields.
The dimensionless size of the periodic box $n_L$ is set as $10$ for the evolution of vortex dipole.
The Fourier spectral method is used for the boundary spatial differentiation with discrete pointd $N_x \times N_y=256\times 256$, and the differentiation along $z$ adopt the Chebyshev pseudo-spectral method with the discrete pointd $N_z=32$.
With these setups, the time evolution proceeds after specifying initial conditions.

The stress tensor in the bulk read as
\begin{equation}
\begin{aligned}
\tau^\mu_{\ \nu}&=\frac{\sqrt{-g}}{2}\left(F^{\mu A}F_{\nu A}-\frac{1}{4}\delta^\mu_{\ \nu}F_{AB}F^{AB}+2\textrm{Re}[(D^\mu\Psi)^*D_\nu\Psi]\right.\\
&\left.-\frac{1}{2}\delta^\mu_{\ \nu}(D^A\Psi (D_A\Psi)^*+m^2\Psi\Psi^*)\right),
\end{aligned}
\end{equation}
which satisfies the conservation equation $\partial_i \tau^i_{\ j}=-\partial_z\tau^z_{\ j}$ where $i, j$ represent the boundary coordinates ($t,x,y$). Thus the energy varying at boundary is equal to the flux along the gravity direction $z$, and the dissipation rate at boundary is given by the energy flux
\begin{equation}
\tau^{z}_{\ t}|_{\text{horizon}}=\frac{1}{2}\left(F^{z \mu}F_{t \mu}+2\textrm{Re}[(D^z\Psi)^*D_t\Psi]\right)|_{z=1}.
\end{equation}

The compressible kinetic energy $\mathcal{E}_c= |\textbf{V}|^2$. $\textbf{V}=(V_x, V_y)$ is the compressible velocity, which is defined as
$\text{iFFT}[\textbf{k}(\textbf{k} \cdot \text{FFT}[\mathcal{V}])/|\textbf{k}|^2]$ where $\textbf{k}=(k_x, k_y)$ is the wave vector. $\mathcal{V}=(\sqrt{\rho}v_x,\sqrt{\rho}v_y)$ where $\rho=|\psi|^2$ is the condensate density and $(v_x,v_y)$ is the superfluids velocity.

\bibliography{ref}

\end{document}